\documentclass[a4paper]{jpconf}
\usepackage{graphicx}
\usepackage[center]{subfigure}
\usepackage{xspace}
\usepackage{footnote}
\usepackage{amsmath}
\usepackage{amssymb}
\usepackage{cite}   

\newlength{\lenerrbars}
\newlength{\lencontours}
\newlength{\lencls}
\begin{document}
\title{A general approach for testing non-cold dark matter at small cosmological scales}

\author{R Murgia}

\address{SISSA and INFN, Sezione di Trieste, via Bonomea 265, 34136 Trieste, Italy}

\ead{riccardo.murgia@sissa.it}

\begin{abstract}
We present a general approach for modelling the small-scale suppression in the linear matter power spectrum induced by the presence of non-cold dark matter.
We show that the new parametrisation accurately describes a large variety of non-thermal scenarios, removing the need to individually
test each of them. We discuss the first astrophysical constraints on its free parameters and we outline the next steps for pursuing a full statistical data analysis.
\end{abstract}

\section{Introduction}
\label{sec:intro}
According to the standard cosmological model ($\Lambda$CDM model), the universe today
is mainly composed by a cosmological constant ($\Lambda$) and by cold dark matter (CDM). While being in excellent
agreement with cosmic microwave background and large-scale structure observations, this
paradigm shows some tensions with structure formation data at sub-galactic scales,
often denoted as the CDM ``small-scale crisis''. Assuming the standard cosmological model, indeed, $N$-body simulations
predict too many dwarf galaxies within the Milky Way (MW) virial radius (\emph{missing satellite} problem~\cite{Klypin:1999uc,Moore:1999nt}) and too much dark matter (DM) in the innermost regions of galaxies (\emph{cusp-core} problem~\cite{2010AdAst2010E...5D}), with respect to observations.
Furthermore, the dynamical properties of the most massive MW satellites are not correctly predicted by simulations (\emph{too-big-to-fail} problem~\cite{2011MNRAS415L40B,2012MNRAS4221203B}).
These discrepancies may be relaxed either by baryon physics, still not perfectly included in cosmological simulations~\cite{Okamoto:2008sn,Governato:2012fa}, or by
modifying the standard CDM framework, given that the fundamental nature of DM is still unknown~\cite{Schneider:2016ayw}.

DM candidates are generally categorised according to their velocity dispersion, which defines a free-streaming length.
On scales smaller than their free-streaming length, density fluctuations are erased and gravitational
clustering is suppressed. The velocity dispersion of CDM candidates is by definition so small that the corresponding 
free-streaming length does not have any influence on cosmological structure formation.
On the other hand, various non-cold DM (nCDM) scenarios predict structure formation to be suppressed at small cosmological
scales and thus have been studied as a viable solution for the small-scale crisis (e.g., sterile neutrinos~\cite{Adhikari:2016bei,Konig:2016dzg}, ultralight scalars~\cite{Hu:2000ke,Marsh:2013ywa,Hui:2016ltb}, mixed (cold + warm) DM fluids~\cite{Schneider:2016ayw,Viel2005},
Self-Interacting DM (SIDM)~\cite{Cyr-Racine:2015ihg,Vogelsberger:2015gpr}).

\looseness= -1 The suppression in the matter power spectrum induced by nCDM can be characterised by different strength and shape, depending 
on the fundamental nature of the DM candidate.
\looseness= -1 Up to now, lots of efforts have gone into exploring the astrophysical consequences of
thermal Warm DM (WDM) models, i.e.~candidates with a Fermi-Dirac or Bose-Einstein momentum distribution, which
implies a very specific shape of the small-scale suppression, only depending on the WDM particle mass~\cite{Viel:2013apy,Irsic:2017ixq}.
\looseness= -1 However, most of the nCDM candidates listed above, well motivated by theoretical particle physics, 
do not feature a thermal momentum distribution: this can yield to non-trivial suppressions in their power spectra, not adequately captured
by the thermal WDM case~\cite{Murgia:2017lwo,Baur:2017stq}.

In this paper we present a general analytical fitting formula which accurately reproduces the small-scale power suppression
induced by the most viable (non-thermal) nCDM scenarios: sterile neutrinos, mixed (cold +
warm) fluids, ultralight scalar DM, as well as other models suggested by effective theory of structure formation (ETHOS).
We also discuss the first, preliminary constraints on its three free parameters from Lyman-$\alpha$ forest data, based on linear theory.
We finally sketch the next steps in order to perform a more comprehensive data analysis.

\section{A new general approach}
\label{sec:approach}

\begin{figure}[b]
\centering
\includegraphics[width=7.5cm]{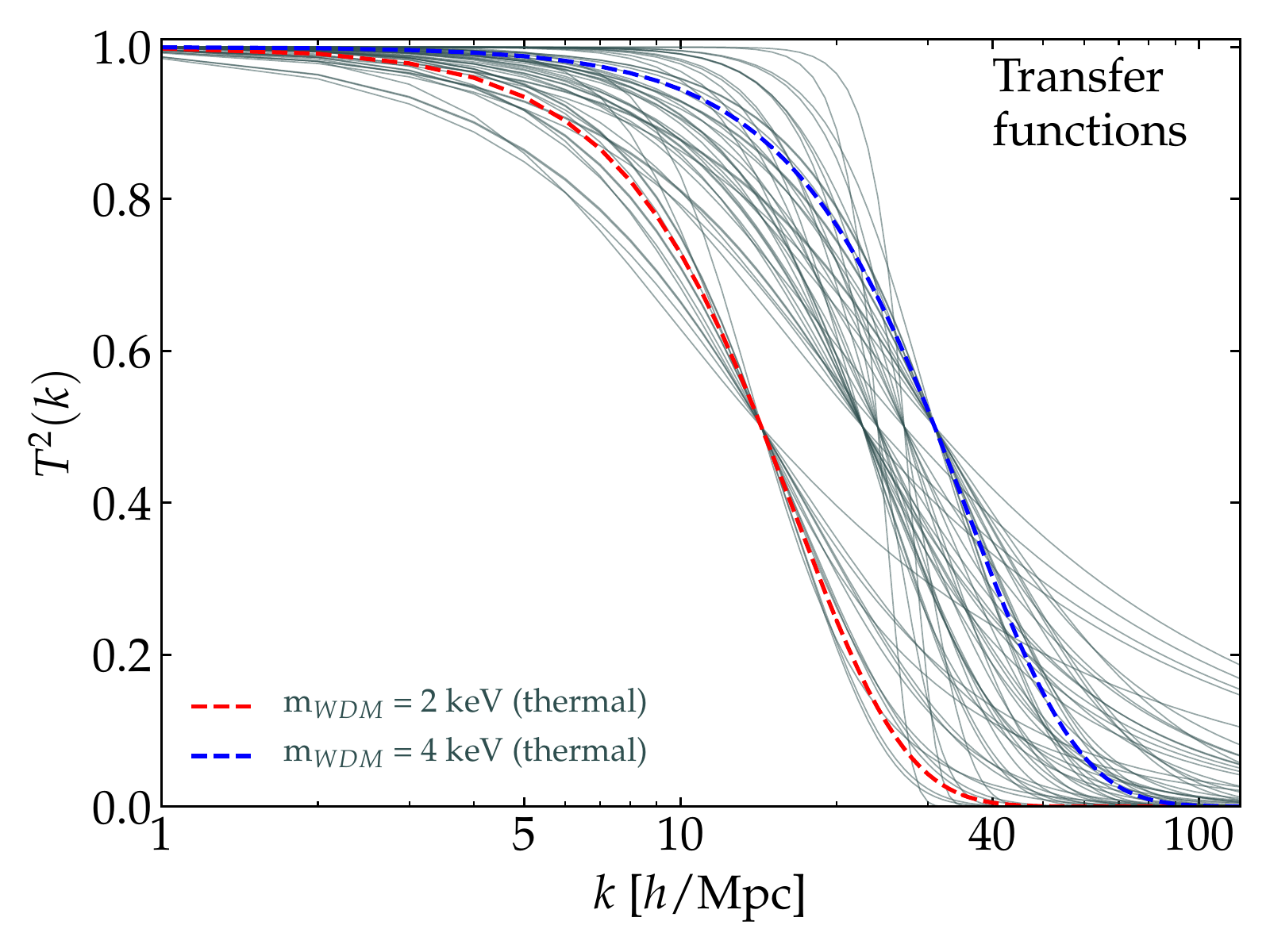}
\caption{\label{fig:Tk}
Here we report the plot of the 55 transfer functions, computed through Eq.~\eqref{eq:Tgen} and associated with the $\{\alpha,\beta,\gamma\}$-combinations considered in the analyses performed by the authors of~\cite{Murgia:2017lwo}. The red and blue dashed lines represent the ``old'' transfer functions (computed via Eq.~\eqref{eq:Viel}) for $m'_x=2 ~\rm{keV}$ and $m'''_x=4 ~\rm{keV}$, respectively.}
\end{figure}

The small-scale suppression of the matter power spectrum $P(k)$, due to the existence of nCDM, is usually 
described by the transfer function $T(k)$. It is defined as follows:
\begin{equation}\label{eq:Tkdef}
 T^2(k) = \left[ \frac{P(k)_{\rm{nCDM}}}{P(k)_{\rm{CDM}}} \right],
\end{equation}
i.e.~the square root of the ratio of the linear power spectrum in the presence of nCDM
with respect to that in the presence of CDM only, for fixed cosmological parameters.
For the particular case of thermal WDM, the transfer function may be approximated by the analytical
fitting function~\cite{Bode2001}
\begin{equation}\label{eq:Viel}
 T(k) = [ 1 + (\alpha k)^{2\mu} ]^{-5/\mu},
\end{equation}
where $\alpha$ is the only free parameter and $\mu = 1.12$.
Therefore, bounds on the mass of the thermal WDM candidate are easily converted into constraints on $\alpha$,
through the following formula~\cite{Viel2005}:
\begin{align}\label{eq:alphaold}
 \alpha = 0.24 \left( \frac{m_x/T_x}{1~\rm{keV}/T_\nu} \right)^{-0.83} \left( \frac{\omega_x}{0.25(0.7)^2} \right)^{-0.16}{\rm{Mpc}}\\
 = 0.049 \left( \frac{m_x}{1~\rm{keV}} \right)^{-1.11} \left( \frac{\Omega_x}{0.25} \right)^{0.11} \left(\frac{h}{0.7}\right)^{1.22}h^{-1}\rm{Mpc}~,
\end{align}
with $m_i$ being the mass, $T_i$ the temperature, $\Omega_i$ the abundance of the $i$-th species and $\omega_i \equiv \Omega_ih^2$. The index $i = x,\nu$ stands for WDM and active neutrinos, respectively.

Let us now define the half-mode scale, $k_{1/2}$, as the wave-number for which $T^2\equiv0.5$,
and introduce the following generalisation of Eq.~\eqref{eq:Viel}~\cite{Murgia:2017lwo}:
\begin{equation}\label{eq:Tgen}
 T(k) = [ 1 + (\alpha k)^{\beta} ]^{\gamma},
\end{equation}
so that $k_{1/2}$ is a function of the three parameters $\alpha$, $\beta$ and $\gamma$,~i.e.
\begin{equation}\label{eq:k12}
 k_{1/2} = ((0.5)^{1/2\gamma}-1)^{1/\beta})\alpha^{-1}.
\end{equation}
Via Eqs.~\eqref{eq:Viel} and \eqref{eq:alphaold} we have a one-to-one correspondence between $m_x$ and
$\alpha$. On the other hand, through Eqs.~\eqref{eq:Tgen} and \eqref{eq:k12}, bounds on the DM mass are mapped to
3D surfaces in the $\{\alpha,\beta,\gamma\}$-space. In other words, given a value of $k_{1/2}$ which corresponds 
to a certain thermal WDM mass, Eq.~\eqref{eq:k12} allows to compute the corresponding surface in a 3D parameter space.

Recent analyses claim that thermal warm DM candidates with masses of the order of 3 keV can induce a suppression in 
the corresponding matter power spectra such that the CDM small-scale crisis vanishes or it is largely reduced~\cite{Lovell:2015psz,Lovell:2016nkp}.
It is thus interesting to explore the volume of the $\{\alpha,\beta,\gamma\}$-space associated to thermal WDM masses between 2 and 4~keV 
by building a 3D grid in the parameter space which samples that volume. Each of the grid points is univocally identified by a certain
$\{\alpha,\beta,\gamma\}$-combination corresponding to a different nCDM model.

In Fig.~\ref{fig:Tk} we plot 55 transfer functions, computed through Eq.~\eqref{eq:Tgen} and associated with the $\{\alpha,\beta,\gamma\}$-combinations considered in the analyses performed by the authors of~\cite{Murgia:2017lwo}.

Let us stress that the position of the half-mode scale $k_{1/2}$ is still set by the value of $\alpha$, even in the new general
parametrisation, while $\beta$ and $\gamma$ are responsible for the shape of the transfer functions before and after $k_{1/2}$, respectively.
$\beta$ has to be greater than zero in order to have meaningful transfer functions, since $\beta < 0$ gives a $T(k)$ that differs from 1 at large scales.
The larger is $\beta$, the flatter is the transfer function before $k_{1/2}$. Analogously, the larger is $|\gamma|$, the sharper is the cut-off on small scales.

Fig.~\ref{fig:Tk} clearly shows that the new fitting formula is flexible enough to disentangle even tiny differences in the shape of the power suppression and, thus, to discriminate between distinct (non-thermal) nCDM models with power spectra suppressed at very similar scales.

\section{The ``area criterion'' for the Lyman-$\alpha$ forest}
\label{sec:lyman}

\begin{figure}[b]
\centering
\includegraphics[width=8cm]{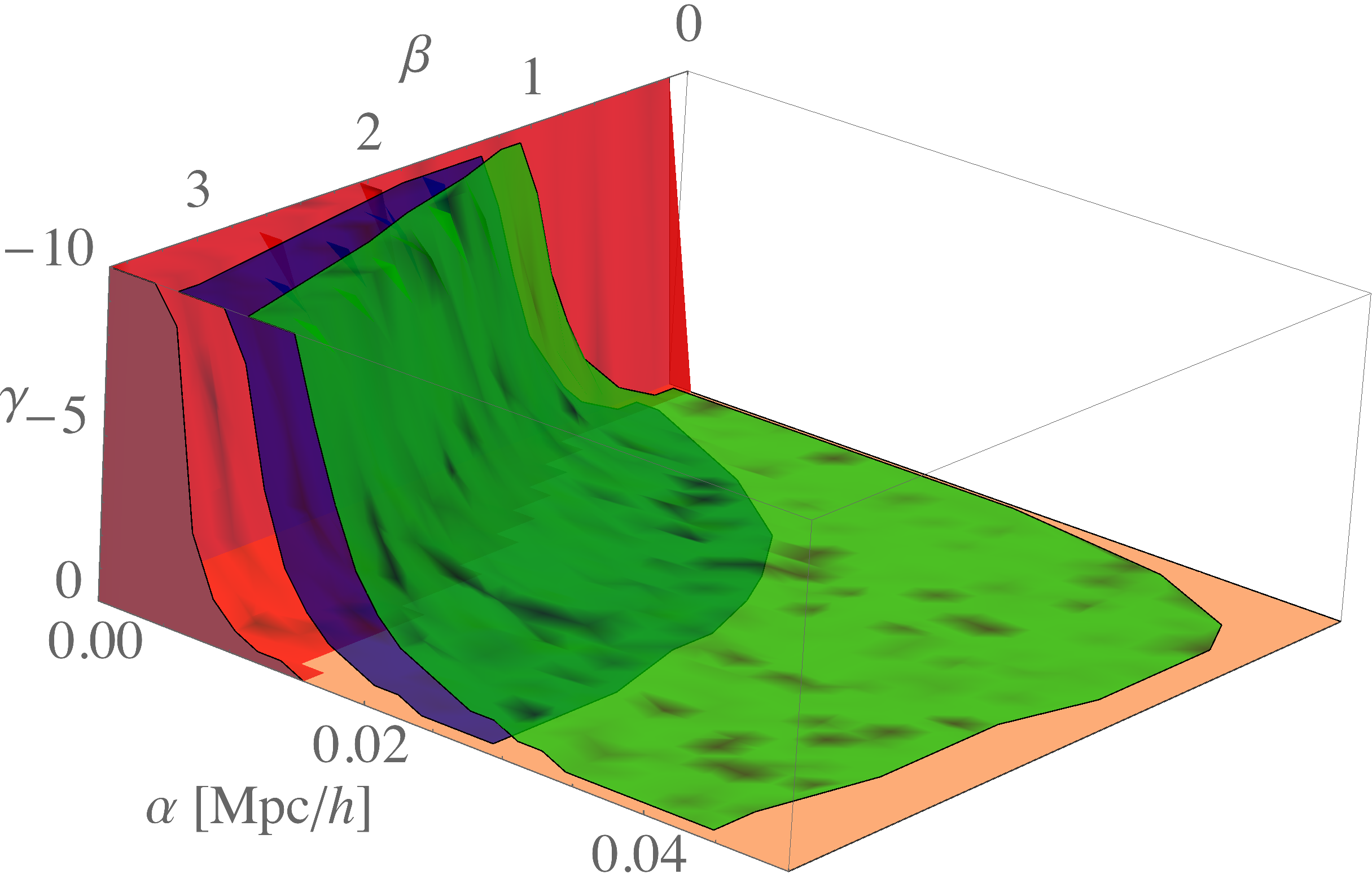}
\caption{\label{fig:abg_contours}
Here we show a 3D contour plot in the $\{\alpha, \beta, \gamma\}$-space (from~\cite{Murgia:2017lwo}) which represents the volume of the parameter space in agreement with Lyman-$\alpha$ forest data, according to the area criterion. The red contour is the 1$\sigma$~C.L.\ limit, while the blue and green contours are the 2$\sigma$ and 3$\sigma$~C.L. constraints, respectively. All those models associated to $\{\alpha, \beta, \gamma\}$-triplets sampling the non-coloured region are thereby excluded at 3$\sigma$~C.L. by our analysis.}
\end{figure}

We now present a simple method, based on linear theory, for testing different DM scenarios with the Lyman-$\alpha$ forest~\cite{McQuinn:2015icp}, which is 
produced by the absorption of the inhomogeneous distribution of the intergalactic neutral hydrogen along different line of sights to distant quasars~\cite{Viel:2001hd}.
Thereby, it constitutes a key observable for efficiently investigating the matter power spectrum at small cosmological scales, namely $0.5~{\rm Mpc}/h \lesssim  \lambda \lesssim 100~{\rm Mpc}/h$~\cite{Viel:2013apy,Irsic:2017ixq}.

Absolute constraints on DM properties can be extracted from Lyman-$\alpha$ forest data only through a comprehensive statistical analysis. However, it is possible to look into deviations with respect to the most updated limits on the mass of thermal WDM candidates, obtained with a recent thorough analysis~\cite{Irsic:2017ixq}. This can be done by applying the ``area criterion'' introduced in~\cite{Murgia:2017lwo}, which we briefly summarise in the next paragraphs.

\looseness=-1 The deviation of a model with respect to the standard CDM case is parameterised by the ratio
\begin{equation}\label{eq:rk}
 r(k) = \frac{P_{1\rm{D}}(k)}{P^{\rm{CDM}}_{1\rm{D}}(k)},
\end{equation}
where $P_{1\rm{D}}(k)$ is the 1D power spectrum of the model that we are considering, computed by the following integral on the 3D linear matter power spectrum, $P(k')$, at redshift $z=0$:
\begin{equation}\label{eq:pk1d}
 P_{1\rm{D}}(k)=\frac{1}{2\pi} \int\limits_k^\infty {\rm d}k'k'P(k').
\end{equation}
In order to find out whether a model deviates more or less from the standard CDM case, with respect to the reference model that we have chosen, we adopt the following criterion: a model is rejected if it shows a larger power suppression with respect to the reference one. The suppression in the power spectra is computed via the following estimator:
\begin{equation}\label{eq:deltaA}
 \delta A \equiv \frac{A_{\rm CDM} - A}{A_{\rm CDM}},
\end{equation}
\looseness=-1 where $A$ is the integral of $r(k)$ over the range of scales probed by Lyman-$\alpha$ observations ($0.5~h/{\rm Mpc} < k < 20~h/{\rm Mpc}$ for the MIKE/HIRES+XQ-100 dataset used in~\cite{Irsic:2017ixq}), i.e.
\begin{equation}\label{eq:A}
A = \int\limits_{k_{\rm min}}^{k_{\rm max}} {\rm d}k\ r(k),
\end{equation}
so that $A_{\rm CDM} \equiv k_{\rm max} - k_{\rm min}$, by definition.

\looseness=-1 The area criterion that we have sketched above can now be applied for constraining the $\{\alpha,\beta,\gamma\}$ parameter space.
\looseness=-1 We take as reference $m_{\rm WDM} = 5.3$~keV (2$\sigma$~C.L.), 
which is the most stringent lower limit on thermal WDM masses from Lyman-$\alpha$ forest data up to date~\cite{Irsic:2017ixq}.
\looseness=-1 By plugging the linear power spectrum associated to this reference model into Eqs.~\eqref{eq:rk} and~\eqref{eq:deltaA}, we find $\delta A_{\rm REF} = 0.21$, which is the estimate of the small-scale power suppression for nCDM models that are excluded at $2\sigma$~C.L. by Lyman-$\alpha$ forest data. \looseness=-1 Analogously, we compute the same estimator $\delta A$ in a grid in the $\{\alpha, \beta, \gamma\}$-space, where each grid point corresponds to a distinct nCDM model, and we accept (at 2$\sigma$ C.L.) only those $\{\alpha, \beta, \gamma\}$-combinations that display a suppression $\lesssim 21\%$ with respect to the CDM power spectrum (i.e.~those models for which $\delta A < \delta A_{\rm REF}$).

In Fig.~\ref{fig:abg_contours} we show a 3D contour plot, which represents the volume of the $\{\alpha, \beta, \gamma\}$-space that contains models in agreement with Lyman-$\alpha$ forest data, according to the area criterion.

\section{Connection with theoretical particle physics models}
\label{sec:tpp}
We now aim to highlight the link between the fundamental nature of DM and our 3D parametrisation. To do so, one should compare the predictions in terms of 
structure formation in order to determine to which extent the 3-parameter fitting formula matches the ``true'' transfer functions associated with various viable nCDM scenarios provided by theoretical particle physics.
By doing this, we can figure out whether the $\{\alpha, \beta, \gamma\}$-fit to a certain nCDM model leads us to the same conclusion about its validity when confronted with the Lyman-$\alpha$ forest constraints, computed by applying the area criterion.

The authors of~\cite{Murgia:2017lwo} have performed this analysis by considering most of the known nCDM particle scenarios:
resonantly produced (RP) sterile neutrinos, sterile neutrinos by particle decays, cold + warm DM fluids, ultralight scalar DM and ETHOS models. 
For each class of models, 5 different $\{\alpha, \beta, \gamma\}$-combinations corresponding to different properties of the given scenario (e.g.~different DM masses, lepton asymmetry, nCDM abundance) have been selected and compared with the analogous actual transfer functions.
The main result is that the $\{\alpha, \beta, \gamma\}$-fit predictions depart from the ``true'' results by few per cent at most, so that the fitted transfer functions practically always lead to the same conclusion provided by the actual models.

Therefore, whenever one wants to test a nCDM particle scenario with structure formation data, it is sufficient to match the resulting transfer functions to Eq.~\eqref{eq:Tgen} and check whether the fitted 3D points are allowed, i.e.~whether they sample the coloured volume of the parameter space shown in Fig.~\ref{fig:abg_contours}. However, being the results reported in~\cite{Murgia:2017lwo} based on linear theory only, they must be considered as a first step in the direction of a more comprehensive analysis. This analysis will be based on a large suite of high resolution hydrodynamical simulations, in order to extract more accurate constraints from Lyman-$\alpha$ forest data. In the next section we briefly outline the path towards such extensive study.

\section{Towards a full Monte Carlo Markov Chain analysis}
\label{sec:mcmc}
The physical observable for Lyman-$\alpha$ forest experiments is the flux power spectrum, $P_{\rm{F}}(k,z)$, rather than the 1D (or 3D) linear matter power spectrum. Nevertheless, two distinctive characteristics of the Lyman-$\alpha$ physics suggest that the area criterion analysis could be also quantitatively correct. Firstly, the use of Eq.~\eqref{eq:rk} with flux power spectra is justified by the existing relation between matter and flux power spectra, namely $P_{\rm{F}}=b^2(k)P_{3\rm{D}}(k)$, where the bias factor $b^2(k)$ differs very little for models reasonably close to the standard CDM scenario~\cite{Viel2005}. Moreover the area criterion is motivated by the fact that the peculiar velocities of the intergalactic medium (generally $<100$ km/s) tend to spread the small-scale power within a relatively broad range of wave-numbers in the explored region~\cite{gnedin02}.

\looseness=-1 It is thus worth to test the accuracy of the area criterion against a full statistical data analysis.
\looseness=-1 This has been done for the first time in~\cite{takeshi}, in the context of ultralight scalar DM, i.e.~in a 2D parameter space where different models are identified by different combinations of scalar DM mass and abundance $\{m,F\}$.
\looseness=-1 The authors of~\cite{takeshi} have performed both the area criterion investigation and a full Monte Carlo Markov Chain (MCMC) analysis of the $\{m,F\}$-space, finding agreement at $10\%$ level between the 2$\sigma$ contours resulting from the two different methods.

Therefore, given that the area criterion represents a simple and intuitive, yet approximate, method for constraining the parameter space of nCDM models, a more precise and extensive approach, not limited to linear theory, is needed. Indeed, we are currently developing a MCMC code which samples the $\{\alpha, \beta, \gamma\}$-volume enclosed by a non-regular grid of 55 full hydrodynamical simulations, each of them associated to a distinct combination of the three parameters, in order to compare the corresponding flux power spectra with the actual physical observables provided by Lyman-$\alpha$ forest experiments. This will allow to determine accurate absolute bounds on the three parameters of Eq.~\eqref{eq:Tgen}.

\section{Conclusions}
\label{sec:conclusions}
In spite of strong efforts both in particle physics and cosmology, the fundamental nature and composition of DM remain undiscovered.
Nevertheless, the majority of the constraints on nCDM properties currently available only apply to the very specific shape of the power suppression corresponding to the thermal WDM case.

Given that most of the viable DM candidates provided by theoretical particle physics are not adequately described by the oversimplified concept of thermal WDM, we have presented a new analytical fitting formula for the transfer function $T(k)$~\cite{Murgia:2017lwo}. Due to the mutual dependence among its three free parameters, the new general formula is capable to reproduce a huge variety of shapes in the suppression of the linear matter power spectrum.
We have pointed out that the new parametrisation is able to embrace all the most viable nCDM particle scenarios,
such as sterile neutrinos, cold + warm DM fluids, ultralight scalars and DM models from effective theory of structure formation (ETHOS).
We have discussed the first constraints on the three free parameters characterising the new approach, obtained by applying an approximate yet very intuitive 
method, based on linear theory, i.e.~the area criterion for Lyman-$\alpha$ forest data~\cite{Murgia:2017lwo, takeshi}.
The Lyman-$\alpha$ forest constitutes, in fact, a powerful tool for constraining DM properties down to very small cosmological scales.

The results discussed here represent a first step towards a fully general modelling of the small-scale departures from the standard CDM model,
which will include an accurate and extensive MCMC analysis of Lyman-$\alpha$ forest data, providing absolute limits easily translatable to bounds on the fundamental nCDM properties through the scheme that we have illustrated.

\section*{References}

\bibliographystyle{unsrt2}
\bibliography{wdm}

\end{document}